# Deciphering general characteristics of residues constituting allosteric communication paths

Girik Malik[1], Anirban Banerji[1], Maksim Kouza[1,3], Irina A. Buhimschi[2,4], Andrzej Kloczkowski*[1,4]

[1] Battelle Center for Mathematical Medicine, The Research Institute at Nationwide Children's Hospital, Columbus, OH, USA
[2] Center for Perinatal Research, The Research Institute at Nationwide Children's Hospital, Columbus, OH, USA
[3] Faculty of Chemistry, University of Warsaw, Pasteura 1, 02-093 Warsaw, Poland
[4] Department of Pediatrics, The Ohio State University College of Medicine, Columbus, OH, 43215, USA

*Email: Andrzej.Kloczkowski@nationwidechildrens.org

**Abstract:**
Considering all the PDB annotated allosteric proteins (from ASD - AlloSteric Database) belonging to four different classes (kinases, nuclear receptors, peptidases and transcription factors), this work has attempted to decipher certain consistent patterns present in the residues constituting the allosteric communication sub-system (ACSS). The thermal fluctuations of hydrophobic residues in ACSSs were found to be significantly higher than those present in the non-ACSS part of the same proteins, while polar residues showed the opposite trend.

The basic residues and hydroxyl residues were found to be slightly more predominant than the acidic residues and amide residues in ACSSs, hydrophobic residues were found extremely frequently in kinase ACSSs. Despite having different sequences and different lengths of ACSS, they were found to be structurally quite similar to each other – suggesting a preferred structural template for communication. ACSS structures recorded low RMSD and high Akaike Information Criterion(AIC) scores among themselves. While the ACSS networks for all the groups of allosteric proteins showed low degree centrality and closeness centrality, the betweenness centrality magnitudes revealed nonuniform behavior. Though cliques and communities could be identified within the ACSS, maximal-common-subgraph considering all the ACSS could not be generated, primarily due to the diversity in the dataset. Barring one particular case, the entire ACSS for any class of allosteric proteins did not demonstrate "small world" behavior, though the sub-graphs of the ACSSs, in certain cases, were found to form small-world networks.

# 1. Introduction:

Starting from Monod−Wyman−Changeux(1) and Koshland−Némethy−Filmer(2) models, investigations of allosteric regulation of protein function have over half-a-century long, rich and multifaceted history. There are so many excellent reviews that have attempted to capture the essence of various aspects of research on this topic that even a cursory enlisting of them will run for pages. The most recent ones include(3-8). To summarize these efforts, one can merely observe that while a lot has been unearthed about the physico-chemical nature of allosteric signal transduction, the various modes through which the long-distant communication is achieved, the structural details of cooperativity revealed during this process, there are still significant aspects of allosteric regulation, especially in the context of generalized characterization of the process, that need to be better understood. The present work reports a few generalized findings about the allosteric communication.

Regardless of the approach adopted to study allostery, we know that large-scale conformational transitions in a protein triggered by a chemical event taking place at a distant part of that same protein, say, binding of a ligand, is one of the signatures of allostery demonstrated by many allosteric systems(9). While this view of studying allostery as a binding event involving conformational changes - all involving a single propagation pathway(10), is one of the major approaches to investigate allostery, many other schools exist. There are four principal types of approaches (though, they are not entirely non-overlapping); viz., a: the evolution-based sequence alignment-centric investigations to identify the probable allosteric communication paths(11-14); b: studying the coupling synchronizations between distant locations in protein space by investigating residue-residue contacts (15-17); c: the MD-simulation based approach to decipher the process of coupled conformational changes and directional perturbation(18-22); d: works based on 'elastic network models' and normal mode analyses that, building upon the observation that distant residues having highly correlated fluctuations may form the parts of allosteric communication, allow computations of correlations among instantaneous fluctuations of different residues in proteins (23-25). It is important to state that we neither claim that the above classification is complete or presents a picture of lineages of how the aforementioned approaches were derived from the past research (though, readers may benefit from(26) for a detailed exposition on both counts). What transpires from these largely case-specific studies, however, is simple; though a huge bulk of information about allostery has been compiled from various perspectives over many years, general methods are needed now to quantitatively describe various aspects of allosteric communication paths or to predict them. But then, before attempting such works, it will probably be beneficial if at least a (somewhat) generalized characterization of allosteric communication paths was available. Because the allosteric communication paths are constituted by a certain subset of residues, we attempted in the present work, to provide a quantifiable difference between the residues involved in allosteric communications and those which are not involved. Because of that, our study revolved principally around identifying the statistical and graph-theoretical differences between the two aforementioned set of residues. We started by noticing that the sheer diversity of cases demonstrating allosteric communications prevents an assumption-based construction of any mathematical framework, rather such a framework would have to emerge from available data. For example, while the oldest views of allostery(1) (2)were rooted on the multisubunit systems, an almost equally old report can be found that describes allosteric transition for monomeric

enzymes(27). Similarly, while the necessary role of the symmetrical multimeric proteins have been discussed in one report(28), roles of monomeric proteins or that of multimeric proteins apparently without symmetry in achieving allosteric communications have been reported also(29, 30). Taking notice of these differences, we chose to look beyond the symmetry aspects, be it crystallographic or self-similar(31), but by resorting to a collection of related approaches to decipher some consistent patterns embedded latently in structural, biophysical and topological nature of allosteric communication sub-structures(ACSS).

We focused on the analysis of mobilities of residues forming ACSS. We compared protein fluctuations derived from crystallographic Debye-Waller B-factors of experimentally solved crystal structures with those obtained from the root mean square fluctuations (RMSF) profile from computational modelling.

## 2.1. Materials :

The curated database ASD(Allosteric Database)(32) was used to retrieve protein structures with information about the identified allosteric communication paths. In some cases we spotted differences in the description of protein structures provided by the ASD and the PDB. These cases were not considered for the study. Retaining the typification scheme provided by the ASD, the finally selected set of 30 proteins that were further divided in four groups: the kinases, the nuclear receptors, the peptidases and the transcription factors. The PDB IDs of these 30 proteins are: 1CZA, 1DKU, 1E0T, 1PFK, 1S9I, 1SQ5, 2BTZ, 2JJX, 2OI2, 2VTT, 2XRW, 3BQC, 3EQC, 3F9M, 3MK6, 4AW0 (- as the kinase group of allosteric proteins); 1IE9, 1XNX, 2AX6 3S79 (- group of nuclear receptors); 1SC3, 2QL9, 4AF8 (- the peptidases); and 1JYE, 1Q5Y, 1R1U, 1XXA, 2HH7, 2HSG, 3GZ5 (- the transcription factors).

## 2.2. Methodology:

Resorting to a coarse-grained representation of residues, and a reduced amino acid alphabet is more likely to lead to generalized ideas from the investigation of the ACSS of 30 proteins. A mere two-letter hydrophobic-polar classification of the residues would have been too broad to reveal the complexity of the problem. Thus we resorted to a scheme(33, 34) which has been found to be extremely successful in protein structure prediction studies(35, 36). Here the 20 amino acids are expressed with a reduced 8-letter alphabet scheme; that is: GLU and ASP - as acidic, ARG, LYS and HIS - as basic, GLN and ASN - as amides, SER and THR - as hydroxyls, TRP, TYR, PHE, MET, LEU, ILE and VAL - as hydrophobic, and GLY and ALA - small residues. PRO and CYS are special among the 20 amino acids for obvious reasons; because of their special status, each one of them are placed as singleton groups. This coarse-grained description was used to study both population characteristics of the ACSS constituents and to undertake the network-based investigations of ACSS.

Because the central theme of this study was to measure the extent by which the residues involved in allosteric communication differ from all remaining residues, various tests were conducted throughout the study to compare residues constituting ACSS with non-ACSS ones. Atoms of the residues not identified (viz., color-coded) by the ASD as part of allosteric communication paths, were considered to be non-ACSS residues and atoms.

We were interested to know whether the sub-structures of the allosteric communication paths have structural similarities among themselves, or not; so i.e. the kinase allosteric communication paths will be characterized by a certain set of canonical parameters, while the nuclear receptor's allosteric communication paths will be different by certain (structural) degrees, etc. However, we realized that a mere structural superposition of the ACSS, will fall short of our goal because the sub-structures in question are characterized by different sequence compositions, different lengths, and additionally may contain various gaps and insertions. Because neglecting certain parts of these sub-structures would have caused loss of valuable information, we used THESEUS 2.0 software(37) that superposes multiple protein structures without throwing away the gaps in them. For network-based studies, Python's NetworkX was used as the graphing library, while matplotlib was used for image generation.

Because statistical tests are necessary to categorically establish the general traits in the allosteric proteins and yet, because the present study considers a limited set of allosteric proteins as belonging to different four classes two non-parametric tests (Wilcoxon signed-rank test and Friedman's non-parametric test)(38-40) were employed to ascertain the traits of the obtained results.

In investigating the "small world network" characteristics, methodologies elaborated in (41) were implemented by us; details about the theoretical basis of the methodology, thus, can be found there. To gather the answer to the question of whether or not the ACSSs are SWNs or not, at multiple resolutions, we studied the problem by generating the Erdös-Rényi (E–R) random graph at three probabilities: 0.3, 0.5 and 0.7.

The computational modelling is a key to solving many fundamental problems of molecular biology. Prediction of protein structures and interactions(42) as well as structural transformations taking places during unfolding, folding and aggregation processes have been studied by computer simulations at different levels of resolution and timescales (43-52). For more efficient simulations one uses coarse-grained (CG) models which reduce the complexity of each amino acid by representing it by a single node or group of pseudo atoms (42, 53-55). In our work, the CABS-flex method (56) is used for predicting protein fluctuations. CABS-flex employs a coarse grained CABS model (54) - efficient and versatile tool for modelling protein structure, dynamics and interactions(57-60). In CABS model one uses coarse-grained representation for a protein in which four atoms per residue are retained. The four atoms include $C\alpha$, $C\beta$, center of mass of the rest of side chain and the center of the virtual $C\alpha$- $C\alpha$ bond. A high resolution lattice model with realistic interactions between atoms is implemented at which $C\alpha$ atoms of the polypeptide chain are confined to the simple cubic lattice model with the lattice spacing of 0.61Å. Large number (800) of the possible virtual $C\alpha$-$C\alpha$ bonds orientations keeps the protein conformation space flexible and ensures overcoming lattice anisotropy problems. Unlike the alpha Carbon atoms, which are restricted to the lattice, the remaining atoms in CABS model are off-lattice. Conformations obtained by CABS-flex simulations further can be reconstructed to physically sound atomistic systems using coarse-grained to atomistic mapping methods(57, 61).The interactions between atoms are described by a realistic knowledge-based potential, while protein-solvent interactions are approximated using implicit solvent model. More detailed description of the CABS force field can be found elsewhere (54, 62).

Discrete conformation space, coarse-grained representation and solvent treated in implicit fashion involved in CABS model greatly reduce the tremendous number of degrees of freedom and free energy landscape roughness. Such simplifications seem to be a reasonable compromise compared to more computationally demanding (and in many cases prohibitive due to the huge system size involved) explicit-solvent all-atom modelling. It is worth noting that despite rather drastic above mentioned simplifications and approximations involved in CABS modelling, the results on structure prediction and dynamics from CABS model simulations reasonably agree with both more sophisticated all-atom MD simulations as well as experimental results(57, 60, 63-65).

To address the question whether B-factors from protein data bank file are consistent with mean square fluctuations of atoms from simulations, we perform near-native simulations for three conceptually different proteins, 1Q5Y from transcription factors group, 1SC3 from the peptidases and 2JXX from the kinase group of allosteric proteins.

### 3. Results and Discussions:

#### 3.1: The thermal fluctuation of residues in allosteric communication paths:

Mass distribution in proteins is known to be inhomogeneous (66); thus, it is not realistic to expect that protein atoms populating diverse spatial zones of a protein will be restricted by equal degrees in their allowed structural thermal fluctuations. Alongside the dynamics needed to ensure the propagation of the structural signal through protein, the ACSS residues possess their inherent thermal fluctuational dynamic. The construction of the residual-interaction networks depends on the value of the cutoff distance, that may be different than the commonly used value of 6.5 Å (67). To quantify the extent of fluctuations of the ACSS residues, versus fluctuations of non-ACSS residues, we extracted B-factors from the coordinate files of the protein structures in protein data bank (PDB) (68) for all 30 proteins. Results found from this investigation, however, are far from simple. While the (naive) expectation was to observe that the residues in ACSS have more inherent fluctuations, obtained results demonstrated that the thermal fluctuations are not only residue-specific but also specific to the type of allosteric proteins considered; Table 1 contains the details of this investigation. The results shown in Table 1 reveal certain interesting findings. For example, the hydrophobic residues (barring PHE) in ACSS have been found to represent large thermal fluctuations, significantly larger than the non-ACSS hydrophobic residues. The global averages of B-factors for residues VAL, LEU, ILE and MET in ACSS's have shown a consistent difference (of ~+9.0) in comparison to the non-ACSS ones for the same set of proteins. On the other hand, the polar residues belonging to ACSS have smaller B-factors in comparison to the same non-ACSS residues, barring the case of GLU which has the same B-factors in both cases. However, the differences in B-factors of the polar residues in ACSS and non-ACSS regions are not found to be as high as that observed for the hydrophobic residues. Differences of B-factors in PRO, CYS and GLY were not considered because they occurred in very low number cases (≤ 3).

To assess whether and by what extent the B-factors of different families of allosteric proteins differ from each other, we subjected the mean values to Friedman's non-parametric test (alternatively referred to as 'non-parametric randomized block analysis of variance')(38, 39). We chose to employ Friedman's test because, ANOVA requires the assumptions of a normal distribution and equal variances (of the residuals) to hold, none of which is found to be existing in our case (viz., that in Table 1), while Friedman test is free from the aforementioned restrictions. The null hypothesis for the test was that the B-factors of the four types of ACSS are the same across repeated measures. Result obtained from the test categorically demonstrates that there indeed exists a substantial difference in the B-factors of these four classes of ACSSs. Results obtained from B-factors of four types of ACSS was Freidman $X^2 = 20.4 > 16.266$ (P value at 0.001, with 3 degrees of freedom), whereby the null hypothesis was rejected comprehensively.

To ascertain the degree to which the B-factors of ACSS residues in each of the four classes of allosteric proteins differ from the B-factors of the non-ACSS residues, each of the classes were subjected to Wilcoxon signed rank test (36), which is a non-parametric analogue of paired t-test for correlated samples, without assuming that the population is normally distributed. The null hypothesis for each of the comparisons was that the median difference between pairs of observations is zero. Result obtained from the tests revealed that the B-factors of ACSS residues in each of the classes differed significantly than the B-factors of the non-ACSS residues. For the kinase class of allosteric proteins we found, $W_{kinase}=87 >> 23$ ($[W(\alpha=0.01,17)=23]$); for the peptidase class $W_{peptidase}=8 > 2$ ($[W(\alpha=0.05,7)=2]$ (we note that W is not defined in 0.01 at degrees of freedom 7 (though $W(0.01,8)=0$), whereby, the critical value comparison is being reported at the weaker 0.05 level); for the Nuclear Receptors, $W_{NR}=15 > 5$ ( $[W(\alpha=0.01,11)=5]$); and for the transcription factors, $W_{TF}= 6 > 5$ ($[W(\alpha=0.01,11)=5]$). Thus, the null hypothesis was rejected in each of the four cases with extremely high confidence.

**3.2. Composition of the ACSS population:**

Allosteric signalling achieved at the structural level show certain differences for various proteins(69-71). Thus, we expect to observe differences in composition of ACSS residues for four different classes of proteins. We found that basic residues are more frequent in ACSS than the acidic ones. To demonstrate this prevalence let us take a closer look at the composition of ACSSs for kinases: the acidic residues were found in 12/133 cases, while the basic residues occurred in 32/133 cases. For the class of transcription factors the basic residues in ACSSs occurred in 13/35 cases, whereas the acidic residues occurred in 5/35 cases. The hydrophobic residues were found to occur in ACSSs of kinases with significant frequency (59/133 cases), but were be notably small in ACSS of transcription factors (2/35) and in peptidases (1/11).

Hydroxyl residues were found to be more common in ACSSs than the amide residues, for kinase ACSS: amide residues 6/133, and hydroxyl residues 14/133 cases. PRO and CYS populations although are extremely small in ACSS, show that CYS occurs slightly more frequently than PRO. The small amino acids (GLY and ALA) were found in very small frequency in ACSSs, while TRP was not found as part of any of the ACSSs.

**3.3: Structural Superimposition of Multiple Allosteric Communication Paths:**

Results obtained from the structural superimposition of multiple ACSSs demonstrated clearly that the allosteric communication paths, for any type of allosteric protein, match closely each other in their structures. Upon superimposing the PDB-coordinates of ACSSs of all proteins for each of the four classes using 'Theseus' software we obtained results shown in detail in Supplementary Material:1. Here we report the two most prominent results, a: RMSD for the superposition, and b: the Akaike Information Criterion (AIC). AIC proposed by Akaike(72) has become commonly used tool for statistical comparison of multiple theoretical models characterized by different numbers of parameters. Because the RMSD of two superposed structures indicates their divergence from one another a small value is interpreted as a good superposition. In contrast, the higher magnitude of AIC indicates better superposition.

We found that the ACSS paths, despite belonging to different proteins and corresponding to sequences of varying lengths, consistently demonstrated lower RMSD values and significantly higher AIC values in comparison to non-ACSSs parts of the structures (see the details in the Supplementary Material:1).

**3.4. Network analyses of allosteric communication paths:**

**3.4.1: Centrality of ACSS:**

The process of allosteric signal communication is directional, but the richness of the constructs available to study networks becomes apparent by using non-directional graph-theoretical framework. Thus, instead of asking 'what is the route of allosteric signal propagation for a specific protein?', which is already provided by ADB, we asked questions like: 'how robust the ACSSs are, compared to non-ACSS parts of the proteins?', or , 'how does the fluctuation of one arbitrarily-chosen residue influence the spread of allosteric signal through ACSS?', or , 'how probable is it that allosteric communication occurs through a randomly chosen shortest path between two residues belonging to ACSS?', etc.

To answer these and similar questions of general nature, we started our investigation by studying the centrality aspects of the ACSS network. The centrality metrics quantify the relative importance of a protein residue (viz. the vertex) or a residue-to-residue communication path (viz., an edge) in the network description of ACSS. There are many centrality measures, we chose to concentrate upon three fundamental measures outlined in Freeman's classic works(73, 74), namely: degree centrality, betweenness centrality and closeness centrality. Because our ACSS networks are unweighted, we did not calculate the Katz centrality. Also, though eigenvector centrality (75, 76) and Gould index present refined knowledge of network centrality, they don't basically differ from degree centrality, and were neglected in the present work. Degree centrality reveals the local characteristics of the ACSS graph while the other two measures, (betweenness and closeness centralities), tend to reflect upon the global network structure, because they rely on counting shortest paths.

**3.4.2. Degree Centrality:**

Degree centrality for any protein residue in an ACSS network is calculated in a straightforward way, by counting the number of residue-residue communication links connecting that residue (implementing the classical definition (73) to the context of ACSS). Degree centrality of any ACSS residue provides an idea about the local structure around that residue, by measuring the number of other residues connected to it. We note that degree centrality is a local measure that does not provide any information about the network's global structure. We have found that the average degree centrality of ACSS residues, irrespective of the type of allosteric proteins, is lower than the average degree centrality of the non-ACSS residues. This result, alongside that obtained from the other centrality measures are presented in Table-2. Maintaining equality in the size (i.e. the number of residues considered) in every quartet of networks (one ACSS network and three non-ACSS networks from the same protein), the average degree centrality was calculated for every group of ACSS and non-ACSS residues; which in turn, was calculated after calculating the degree centrality of each residue belonging to ACSS and non-ACSS fragments. These residue-specific details of degree centrality for every protein considered, for ACSS and similar size non-ACSS fragments, are provided in Supplementary Material-2.

We note that the average degree centrality of ACSS fragments consistently show lower values than similar non-ACSS fragments. The kinase and the transcription factor ACSSs have similarly low average values of degree centrality. The nuclear receptors have them significantly higher, while the peptidases ACSSs have the highest average values of degree centrality. We note also that the typical differences between average degree centrality of ACSS and non-ACSS fragments are: ~0.10 for kinases, ~0.8 for nuclear receptors, ~0.9 for peptidases, and ~0.13 for transcription factors.

More importantly, the consistent observation of lower average degree centrality of ACSS fragments irrespective of the type of allosteric proteins suggests that it is a general feature. In terms of the network theory the degree centrality is the diagonal element of adjacency matrix that corresponds to the sum of all off-diagonal elements in a given row/column. In an alternative way, one may as well describe the degree centrality as the number of paths of unit length coming out from a given vertex. Degree centrality shows the ability of a given vertex to influence or to be influenced by the local structure around it. It can be considered as a marker (or even predictor) of immediate effects of propagating through a network. Thus, the consistently lower values of average degree centrality observed in ACSS fragments suggests that nature attempts to shield them from perturbations which may destabilize allosteric communication. In terms of attempts to shield the ACSS from perturbations influencing residual interaction network, degree centrality demonstrates similarity with the concept of eigenvector centrality, the caveat being that the later provides a metric to assess direct and indirect influences taking place over a long time-interval while the former measures immediate effects of a perturbation solely.

### 3.4.3. The global centrality measures:

While the degree centrality provides a measure to assess the possibility of immediate involvement of a residue in influencing the signal communication in residue interaction network of a protein, the concepts of closeness centrality and betweenness centrality provide ideas of how the global topology of the network influences the signal propagation. Closeness centrality of any connected graph measures how "close" a vertex is to other vertices in a network; this is computed by summing up the lengths of the shortest paths between that vertex and other vertices in the network. Closeness of a vertex, thus, can be interpreted as a predictor of how long it may take for that vertex to communicate with all other vertices. In the framework of protein residue connectivity network, the residues with low closeness score can be identified as ones that are separated by short distances from other residues. It can be expected that they receive the structural signal (i.e. instantaneous fluctuation or perturbation) faster, being well-positioned to receive this information early. We indeed found that the average closeness centrality of the ACSS network is lower in comparison to the non-ACSS fragments, for all the types of allosteric proteins. However, the difference between the extent of average closeness centrality between ACSS and non-ACSS fragments was found to vary over a larger scale than what was observed for average degree centrality. For graphs of equal sizes, the difference between the average closeness centrality of ACSS and non-ACSS fragments was found to be: ~0.04 for peptidases, ~0.05 for nuclear receptors, ~0.11 for kinases, and ~0.22, for transcription factors (see Table-2). We also found that while ACSSs of transcription factors and kinases have small values of average closeness centrality, suggesting fast propagation of allosteric signals, nuclear receptors, and peptidases are significantly less capable of rapid transmission of the signal in the allosteric communication. Based on values of closeness centrality of ACCS fragments, peptidases are expected to be twice slower in propagating allosteric signals in comparison to transcription factors and kinases.

To assess the extent to which the centrality measures differ for four different classes of proteins, we subjected them to Friedman non-parametric test. The null hypothesis was that the centrality measures across the four different classes are the same under repeated measures. Since the value of the test statistic $Q=13.67 > 12.838$ (p-value), the null hypothesis was rejected comprehensively.

The betweenness centrality provides more idea about the global network structure; for every vertex of the network the betweenness centrality specifies the fraction of the shortest paths (geodesics) that pass through that vertex. In this sense, such measure assesses the influence that a given vertex (residue) has over the transmission of a structural signal. A residue with large betweenness centrality score can be expected to have a large influence on the allosteric signal propagating through the ACSS network. Results obtained by us shown inTable-2, are found to be less clear to interpretation. While the average betweenness centrality of ACSS networks in peptidases is found to be higher in comparison to non-ACSS fragments, the similar data for kinases, nuclear receptors, and transcription factors show opposite behaviour. These results imply that for three out of four considered classes of allosteric proteins the residues constituting the ACSS net do not fall in the geodesics which communicate the allosteric signal. - Though it seems unexpected at the first, a close scrutiny at the definition of betweenness centrality reveals that this is not only possible but can probably be expected. The implicit assumptions in linking the signal transmission with betweenness centrality are, first, that the information propagation through the network takes place along the geodesics, and second, that the information is an undivided chunk.

While such characterizations can be useful in many cases of network science, the determinism of a biological network may include certain provisions for error handling and fault tolerance mechanisms developed through evolution. Thus, propagation of information (the structural signal in the case of allostery) across ACSS although deterministic may not follow the simplistic geodesic-based route. A geodesic, i.e. the shortest path, is identified as the path through which the two given vertices are connected by the smallest number of intermediate vertices. The mean of the shortest path lengths reflects the expected distance between two connected vertices. While it has been reported that residues constituting the ligand binding sites of proteins have low average of their shortest path lengths(77), the ACSS networks studied in the present work tends to suggest that the allosteric signal communication may not take place along the shortest path. Second, as the various recent works of allostery tend to suggest, the structural fluctuation-based signal propagating across protein, may not always be treated as an undivided entity(78-80). Thus, the average betweenness centrality studies for ACSSs need further investigations based on availability of larger datasets in the future.

**3.5. Cliques and Communities in ACSS:**

Cliques are the complete subgraphs, where every vertex is connected to every other vertex. A clique is considered maximal only if it is not found to be a subgraph of some other clique. Communities are identified through partitioning the set of vertices, whereby each vertex is made a member of one and only one community. Because of their higher order connectivity, the cliques detected in protein structures are considered to indicate regions of higher cohesion (in some cases, rigid modules). Do the ACSSs embody certain common characteristics in their connectivities which can be revealed through the cliques and communities? To answer this question, we subjected the ACSSs of each of the four classes of proteins to investigation, which implemented (81) and (82) algorithms. We found that indeed the ACSS modules can be partitioned into cliques and communities(83, 84). These results are presented in Fig.-1

**3.6. Maximum common subgraphs to describe the ACSS:**

A maximal common subgraph of a set of graphs is the common subgraph having the maximum number of edges. Many attempts have been made for the last two decades to apply this methodology in protein science(85-87). Finding the maximal common subgraph is a NP-complete problem (88). To solve this difficult problem a backtrack search algorithm proposed by McGregor(89) and a clique detection algorithm of Koch (88), are traditionally used. However, for our ACSSs, some of which are quite large in size, neither McGregor's nor Koch's algorithm was found to be applicable; primarily because of the huge computational costs incurred by the exponential growth of intermediary graphs of varying sizes. Thus, upon generating the subgraphs for each of ACSSs (using Python's NetworkX), we had to resort to the brute-force method to identify the maximum common subgraph for each of the ACSS classes. In some cases, the number of cliques was found to be large; e.g. for 2BTZ (see Supplementary Mat.-3), while in some other cases only one clique was found (e.g. for 2VTT (Res78:hydrophobic—Res76:basic—Res71:basic); or for 2XRW (Res198:hydrophobic—Res230:basic—Res231:hydrophobic)).

Fig-2 presents five maximum common subgraphs composed of 3 vertices and 3 edges, obtained from five subsets (each comprised of 5 to 7 proteins) of kinase ACSSs. The fact that different combination of subsets of kinase ACSSs were found to represent different maximum common subgraphs, point out, why, finding a general maximum common subgraph (composed of more than two nodes and one connecting edge) for the entire set of kinase ACSSs was found to be impossible for this dataset.

**3.7: How frequently do the allosteric communication paths form small world network?**

Investigating whether in general the ACSS residues belonging to the four different classes of allosteric proteins constitute 'small world' networks (SWN) or not is important; because SWNs are more robust to perturbations, and may reflect an evolutionary advantage of such an architecture(90, 91). There are numerous previous works which talk about SWN(92) and the relevance of SWN in investigating the protein structural networks (93-95) and protein-protein interaction networks(96). The SWN (92), constitute a compromise between the regular and the random networks, because on one hand they are characterized by large extent of local clustering of nodes, like in regular networks, and on the other hand they embody smaller path lengths between nodes, something that is distinctive for random networks. Because of the ability to combine these two disparate properties, not surprisingly, it has been shown that networks demonstrating the 'small-world' characteristics tend to describe systems that are characterized by dynamic properties different from those demonstrated by equivalent random or regular networks (92, 96-100). Greene and Higman (95) have shown that protein structure networks which demonstrate both the long-range and short-range interactions exhibit a SWN character; however, the ones demonstrating only the long-range interactions cease to remain SWNs. Additionally, a study of a benchmarked set of 15 pairs of ACSSs with effector and substrate both present in at least one of the two structures demonstrated that the clusters possessing at least one substrate or an effector molecule exhibited SWN characteristics(101). This inspired us to perform a thorough investigation of the entire set of ACSSs for each of the four classes of allosteric communication paths.

We have found that whether ACSSs exhibit SWN nature or not - is a complex problem; while the complete ACSS of a protein may not always demonstrate SWN characteristics, many sub-graphs of non-trivial lengths of the same ACSS reveal SWN character. To elucidate, the ACSSs of all kinases, all peptidases, all transcription factors, and three nuclear receptors (1IE9, 2AX6, 3S79), were found to not demonstrate SWN characteristics. In fact, because residues constituting the ACSS tend typically to not cluster in a particular region of protein structure but are distributed over entire structure, (except of 1XNX), all the ACSSs (mentioned above) were found to be represented by unconnected graphs. Similar disconnected nature of the graphs was revealed by the non-ACSSs fragments.

To get a better understanding of the topology of networks, the unconnected graphs of ACSSs and non-ACSSs fragments were first segmented into a set of sub-graphs of non-trivial lengths ($\geq 3$ nodes); then, shortest paths in the respective subgraphs were calculated.

The graph fragmentation was performed using Python's NetworkX library (with 'matplotlib' library used for image generation). Structural graphs of all ACSS or non-ACSS fragments were found to be fragmented into more than one sub-graph in almost every case (except of nuclear receptor protein 1XNX). To ascertain the nature of clustering characteristics the mean clustering coefficient as well as the mean shortest paths for each of these sets of subgraphs representing ACSS or non-ACSS were calculated (see Table-3).

The SWN character was observed only for ACSS of 1XNX (Fig.-3). However, the subgraphs of ACSSs with non-trivial number of nodes (namely $\geq 3$) demonstrate the SWN character, for 6 cases, with the details provided in Table-4. Changing the cut-off distance from 6.5Å to another value (e.g. 10Å or larger) doesn't change the fundamental behaviour, because, the residues involved in ACSS will be still scattered over the entire protein structure.

## 3.8 CABS-flex simulations

With the help of CABS modelling, we computed the values of root mean square fluctuations RMSF, shown as red curves (on right Y-axis) on middle plots in Figs. 4a, 4b and 4c. The values of B-factors are shown as black curves (on left Y-axis). Although quantitative comparison between B-factors and RMSFs is not possible due to different temperatures and environmental factors used in simulations and experiments, qualitatively the data agree. The most fluctuating protein residues during near-native state simulations result in a series of peaks in RMSF profile (red curve, right X-axis) which correlate with experimentally measured B-factor values (black curve, left Y-axis). Upper and bottom snapshots in Fig. 4 correspond to protein representation coloured by crystallographic B-factor values from PDB and by RMSF values from CABS-flex simulation, respectively.

1Q5Y consists of two alpha-helices and four beta strands. First two peaks in RMSF profile (Fig. 4a) correspond to regions connecting the first alpha helix with its neighbouring beta-strands. The third peak is associated with beta hairpin loop connecting the second and third beta strands. Finally, the fourth and fifth peaks correspond to the regions connecting second helix with its neighbouring beta-strands. The results from B-factor profile are slightly different. Both alpha-helices have high temperature factors values. As follows from B-factor and RMSF profiles (Fig. 4a) each of two broad peaks corresponding to H1 and H2 on B-factor profile splits into two peaks on RMSF profile. This phenomenon possibly occurs as a result of competition between the energy gain and the entropy loss of different secondary structural elements upon temperature increase (102). At low crystallization temperature complexity of structural variations of alpha-helix ensemble of microstates is comparable to the one from less structured connecting loop ensemble of microstates. However, as temperature T increases less structured connecting regions gain more entropy compared to structured alpha helix, as a consequence at room temperature residues from connecting loops are more destabilized having larger RMSF values than those forming alpha-helices. In other words, as T increases, alpha helix destabilizes to a smaller extent, compared to its connecting loops. However overall, protein fluctuations from RMSF profile can be mapped to B-factor values and agree well with experimentally measured X-ray B-factors

Fig. 4b and Fig. 4c show RMSF and B-factor profiles as well as corresponding snapshots for two different proteins, 1SC3 and 2JXX. RMSF and B-factor profiles have multiple peaks which are interpreted as a sign of least stable parts of proteins studied. Similar to 1Q5Y protein, we observe that connecting loops as well as C- and N-terminals are the most fluctuating parts of protein. The overall trend is that the mobility of atoms obtained from simulations are in good agreement with the crystallographic B-factors.

## 4.Conclusion:

The aim of the present work was to decipher some general patterns of residues forming the ACSS of 30 allosteric proteins, and compare them with non-ACSS residues in the same proteins. Our aim was to report the general quantifiable differences between these two (aforementioned) sets of residues and not to study the general mechanism of allosteric communication. By performing the CABS-based simulations of proteins around their native conformations we demonstrated that protein fluctuations depicted by RMSF profiles can be mapped to B-factors and show satisfactory degree of agreement with experimental data.

Our results may benefit the protein engineering community and those studying the general mechanism of allosteric communication or in general, long-distance communication in proteins. The knowledge of the topological invariants of communication paths and the biophysical, biochemical and structural patterns may help in a better understanding of allostery. As many recent papers(103-107) have pointed out, the long-distance communication features within proteins involve several types of non-linear characteristics that may often be dependent on transient fluctuations, making it difficult to arrive at a generalized dynamic picture. However a generalized static picture of the long-distance communication route can be obtained, which may help to better understand such communication schemes, especially those related to allostery. The present work attempted to report such generalized findings. While certain yet-unknown (to the best of our knowledge) patterns regarding the thermal fluctuation profile of ACSS atoms, the structural and topological nature of the ACSS have come to light, incongruities of our findings regarding the extent of betweenness centrality in ACSS network and their small-world nature indicates the need for more focused studies directed at these issues, which in turn, may shed new light on allosteric signal communication. For example, proteins, in general, are fractal objects with known characteristics of trapping energy(108-111). Do the findings on betweenness and on small-world network nature reported in this work indicate the possibility of energy traps in ACSSs? - We plan to probe into many such questions in future.

**Authors contributions:**

Conceived and Designed research: AB, AK, GM; Performed research: GM, AB, MK; Analyzed results: GM, AB, MK; Wrote the paper: AB, IB, AK, GM, MK.

**Acknowledgements:**

AK acknowledges support by the start-up funds from The Research Institute at Nationwide Children's Hospital and the grants from the National Institutes of Health (2R01GM072014-05A1) and the National Science Foundation (MCB 1021785). M. K. acknowledges the Polish Ministry of Science and Higher Education for financial support through ''Mobility Plus'' Program No. 1287/MOB/IV/2015/0.

**5. References:**
1. Monod J, Wyman J, & Changeux JP (1965) On the Nature of Allosteric Transitions: A Plausible Model. *J Mol Biol* 12:88-118.
2. Koshland DE, Jr., Nemethy G, & Filmer D (1966) Comparison of experimental binding data and theoretical models in proteins containing subunits. *Biochemistry* 5(1):365-385.
3. Nussinov R (2016) Introduction to Protein Ensembles and Allostery. *Chem Rev* 116(11):6263-6266.
4. Ribeiro AA & Ortiz V (2016) A Chemical Perspective on Allostery. *Chem Rev* 116(11):6488-6502.
5. Dokholyan NV (2016) Controlling Allosteric Networks in Proteins. *Chem Rev* 116(11):6463-6487.
6. Guo J & Zhou HX (2016) Protein Allostery and Conformational Dynamics. *Chem Rev* 116(11):6503-6515.

7. Papaleo E, *et al*. (2016) The Role of Protein Loops and Linkers in Conformational Dynamics and Allostery. *Chem Rev* 116(11):6391-6423.
8. Wei GH, Xi WH, Nussinov R, & Ma BY (2016) Protein Ensembles: How Does Nature Harness Thermodynamic Fluctuations for Life? The Diverse Functional Roles of Conformational Ensembles in the Cell. *Chemical Reviews* 116(11):6516-6551.
9. Alberts B, *et al*. (2015) Molecular Biology of the Cell, Sixth Edition. *Molecular Biology of the Cell, Sixth Edition*:1-1342.
10. del Sol A, Tsai CJ, Ma BY, & Nussinov R (2009) The Origin of Allosteric Functional Modulation: Multiple Pre-existing Pathways. *Structure* 17(8):1042-1050.
11. Lockless SW & Ranganathan R (1999) Evolutionarily conserved pathways of energetic connectivity in protein families. *Science* 286(5438):295-299.
12. Suel GM, Lockless SW, Wall MA, & Ranganathan R (2003) Evolutionarily conserved networks of residues mediate allosteric communication in proteins. *Nat Struct Biol* 10(1):59-69.
13. Morcos F, *et al*. (2011) Direct-coupling analysis of residue coevolution captures native contacts across many protein families. *P Natl Acad Sci USA* 108(49):E1293-E1301.
14. Li WF, Wolynes PG, & Takada S (2011) Frustration, specific sequence dependence, and nonlinearity in large-amplitude fluctuations of allosteric proteins. *P Natl Acad Sci USA* 108(9):3504-3509.
15. Venkatakrishnan AJ, *et al*. (2013) Molecular signatures of G-protein-coupled receptors. *Nature* 494(7436):185-194.
16. Weinkam P, Pons J, & Sali A (2012) Structure-based model of allostery predicts coupling between distant sites. *P Natl Acad Sci USA* 109(13):4875-4880.
17. Cheng TMK, Lu YE, Vendruscolo M, Lio P, & Blundell TL (2008) Prediction by Graph Theoretic Measures of Structural Effects in Proteins Arising from Non-Synonymous Single Nucleotide Polymorphisms. *Plos Comput Biol* 4(7).
18. Weiss DR & Levitt M (2009) Can Morphing Methods Predict Intermediate Structures? *Journal of Molecular Biology* 385(2):665-674.
19. Atilgan C, Gerek ZN, Ozkan SB, & Atilgan AR (2010) Manipulation of Conformational Change in Proteins by Single-Residue Perturbations. *Biophys J* 99(3):933-943.
20. Gerek ZN & Ozkan SB (2011) Change in Allosteric Network Affects Binding Affinities of PDZ Domains: Analysis through Perturbation Response Scanning. *Plos Comput Biol* 7(10).
21. Dima RI & Thirumalai D (2006) Determination of network of residues that regulate allostery in protein families using sequence analysis. *Protein Sci* 15(2):258-268.
22. Smock RG & Gierasch LM (2009) Sending Signals Dynamically. *Science* 324(5924):198-203.
23. Bahar I, Chennubhotla C, & Tobi D (2007) Intrinsic dynamics of enzymes in the unbound state and, relation to allosteric regulation. *Curr Opin Struc Biol* 17(6):633-640.
24. Yang Z, Majek P, & Bahar I (2009) Allosteric Transitions of Supramolecular Systems Explored by Network Models: Application to Chaperonin GroEL. *Plos Comput Biol* 5(4).

25. Xu CY, Tobi D, & Bahar I (2003) Allosteric changes in protein structure computed by a simple mechanical model: Hemoglobin T <-> R2 transition. *Journal of Molecular Biology* 333(1):153-168.
26. Fenton AW (2008) Allostery: an illustrated definition for the 'second secret of life'. *Trends Biochem Sci* 33(9):420-425.
27. Frieden C (1970) Kinetic Aspects of Regulation of Metabolic Processes - Hysteretic Enzyme Concept. *J Biol Chem* 245(21):5788-&.
28. Changeux JP & Edelstein SJ (2005) Allosteric mechanisms of signal transduction. *Science* 308(5727):1424-1428.
29. Kern D & Zuiderweg ERP (2003) The role of dynamics in allosteric regulation. *Curr Opin Struc Biol* 13(6):748-757.
30. Swain JF & Gierasch LM (2006) The changing landscape of protein allostery. *Curr Opin Struc Biol* 16(1):102-108.
31. Banerji A & Ghosh I (2011) Fractal symmetry of protein interior: what have we learned? *Cell Mol Life Sci* 68(16):2711-2737.
32. Huang ZM, *et al*. (2014) ASD v2.0: updated content and novel features focusing on allosteric regulation. *Nucleic Acids Res* 42(D1):D510-D516.
33. Feng YP, Kloczkowski A, & Jernigan RL (2007) Four-body contact potentials derived from two protein datasets to discriminate native structures from decoys. *Proteins* 68(1):57-66.
34. Feng Y, Jernigan RL, & Kloczkowski A (2008) Orientational distributions of contact clusters in proteins closely resemble those of an icosahedron. *Proteins* 73(3):730-741.
35. Faraggi E & Kloczkowski A (2014) A Global Machine Learning Based Scoring Function for Protein Structure Prediction. *Biophys J* 106(2):656a-657a.
36. Gniewek P, Kolinski A, Kloczkowski A, & Gront D (2014) BioShell-Threading: versatile Monte Carlo package for protein 3D threading. *Bmc Bioinformatics* 15.
37. Theobald DL & Steindel PA (2012) Optimal simultaneous superpositioning of multiple structures with missing data. *Bioinformatics* 28(15):1972-1979.
38. Friedman M (1937) The use of ranks to avoid the assumption of normality implicit in the analysis of variance. *J Am Stat Assoc* 32(200):675-701.
39. Friedman M (1939) The use of ranks to avoid the assumption of normality implicit in the analysis of variance, (vol 32, pg 675, 1937). *J Am Stat Assoc* 34(205):109-109.
40. Wilcoxon F (1945) Individual Comparisons by Ranking Methods. *Biometrics Bull* 1(6):80-83.
41. Humphries MD & Gurney K (2008) Network 'Small-World-Ness': A Quantitative Method for Determining Canonical Network Equivalence. *Plos One* 3(4).
42. Kmiecik S, *et al*. (2016) Coarse-Grained Protein Models and Their Applications. *Chem Rev* 116(14):7898-7936.
43. Sulkowska JI, Kloczkowski A, Sen TZ, Cieplak M, & Jernigan RL (2008) Predicting the order in which contacts are broken during single molecule protein stretching experiments. *Proteins-Structure Function and Bioinformatics* 71(1):45-60.
44. Scheraga HA, Khalili M, & Liwo A (2007) Protein-folding dynamics: Overview of molecular simulation techniques. *Annu Rev Phys Chem* 58:57-83.

45. Nasica-Labouze J, *et al*. (2015) Amyloid beta Protein and Alzheimer's Disease: When Computer Simulations Complement Experimental Studies. *Chem Rev* 115(9):3518-3563.
46. Kouza M, Co NT, Nguyen PH, Kolinski A, & Li MS (2015) Preformed template fluctuations promote fibril formation: Insights from lattice and all-atom models. *Journal of Chemical Physics* 142(14).
47. Kouza M, Banerji A, Kolinski A, Buhimschi IA, & Kloczkowski A (2017) Oligomerization of FVFLM peptides and their ability to inhibit beta amyloid peptides aggregation: consideration as a possible model. *Physical Chemistry Chemical Physics* 19(19):2990-2999.
48. Chwastyk M, *et al*. (2017) Non-local effects of point mutations on the stability of a protein module. *J Chem Phys* 147(10).
49. Zhao Y & Cieplak M (2017) Proteins at air-water and oil-water interfaces in an all-atom model. *Physical Chemistry Chemical Physics* 19(36):25197-25206.
50. Kouza M, Lan PD, Gabovich AM, Kolinski A, & Li MS (2017) Switch from thermal to force-driven pathways of protein refolding. *J Chem Phys* 146(13):135101.
51. Jernigan RL & Kloczkowski A (2007) Packing regularities in biological structures relate to their dynamics. *Protein Folding Protocols*, (Springer), pp 251-276.
52. Feng Y, Yang L, Kloczkowski A, & Jernigan RL (2009) The energy profiles of atomic conformational transition intermediates of adenylate kinase. *Proteins: Structure, Function, and Bioinformatics* 77(3):551-558 %@ 1097-0134.
53. Shakhnovich E (2006) Protein folding thermodynamics and dynamics: where physics, chemistry, and biology meet. *Chem Rev* 106(5):1559-1588.
54. Kolinski A (2004) Protein modeling and structure prediction with a reduced representation. *Acta Biochim Pol* 51(2):349-371.
55. Liwo A, He Y, & Scheraga HA (2011) Coarse-grained force field: general folding theory. *Phys Chem Chem Phys* 13(38):16890-16901.
56. Jamroz M, Kolinski A, & Kmiecik S (2013) CABS-flex: Server for fast simulation of protein structure fluctuations. *Nucleic Acids Res* 41(Web Server issue):W427-431.
57. Kmiecik S, Gront D, Kouza M, & Kolinski A (2012) From coarse-grained to atomic-level characterization of protein dynamics: transition state for the folding of B domain of protein A. *J Phys Chem B* 116(23):7026-7032.
58. Wabik J, Kmiecik S, Gront D, Kouza M, & Kolinski A (2013) Combining coarse-grained protein models with replica-exchange all-atom molecular dynamics. *Int J Mol Sci* 14(5):9893-9905.
59. Blaszczyk M, *et al*. (2016) Modeling of protein-peptide interactions using the CABS-dock web server for binding site search and flexible docking. *Methods* 93:72-83.
60. Jamroz M, Orozco M, Kolinski A, & Kmiecik S (2013) Consistent View of Protein Fluctuations from All-Atom Molecular Dynamics and Coarse-Grained Dynamics with Knowledge-Based Force-Field. *J Chem Theory Comput* 9(1):119-125.
61. Gront D, Kmiecik S, & Kolinski A (2007) Backbone building from quadrilaterals: a fast and accurate algorithm for protein backbone reconstruction from alpha carbon coordinates. *J Comput Chem* 28(9):1593-1597.

62. Jamroz M, Kolinski A, & Kmiecik S (2014) Protocols for Efficient Simulations of Long-Time Protein Dynamics Using Coarse-Grained CABS Model. *Methods Mol Biol* 1137:235-250.
63. Kolinski A & Bujnicki JM (2005) Generalized protein structure prediction based on combination of fold-recognition with de novo folding and evaluation of models. *Proteins* 61 Suppl 7:84-90.
64. Kmiecik S & Kolinski A (2007) Characterization of protein-folding pathways by reduced-space modeling. *Proc Natl Acad Sci U S A* 104(30):12330-12335.
65. Jamroz M, Kolinski A, & Kmiecik S (2014) CABS-flex predictions of protein flexibility compared with NMR ensembles. *Bioinformatics* 30(15):2150-2154.
66. Banerji A & Ghosh I (2009) A new computational model to study mass inhomogeneity and hydrophobicity inhomogeneity in proteins. *Eur Biophys J Biophy* 38(5):577-587.
67. Sun WT & He J (2011) From Isotropic to Anisotropic Side Chain Representations: Comparison of Three Models for Residue Contact Estimation. *Plos One* 6(4).
68. Berman HM, *et al*. (2000) The Protein Data Bank. *Nucleic Acids Research* 28(1):235-242.
69. Banerji A (2013) An attempt to construct a (general) mathematical framework to model biological "context-dependence". *Syst Synth Biol* 7(4):221-227.
70. Tuncbag N, Gursoy A, Nussinov R, & Keskin O (2011) Predicting protein-protein interactions on a proteome scale by matching evolutionary and structural similarities at interfaces using PRISM. *Nat Protoc* 6(9):1341-1354.
71. Ozbabacan SEA, Gursoy A, Keskin O, & Nussinov R (2010) Conformational ensembles, signal transduction and residue hot spots: Application to drug discovery. *Curr Opin Drug Disc* 13(5):527-537.
72. Akaike H (1981) Citation Classic - a New Look at the Statistical-Model Identification. *Cc/Eng Tech Appl Sci* (51):22-22.
73. Freeman LC (1979) Centrality in Social Networks Conceptual Clarification. *Soc Networks* 1(3):215-239.
74. Freeman LC, Borgatti SP, & White DR (1991) Centrality in Valued Graphs - a Measure of Betweenness Based on Network Flow. *Soc Networks* 13(2):141-154.
75. Bonacich P (1987) Power and Centrality - a Family of Measures. *Am J Sociol* 92(5):1170-1182.
76. Bonacich P (1991) Simultaneous Group and Individual Centralities. *Soc Networks* 13(2):155-168.
77. Atilgan AR, Akan P, & Baysal C (2004) Small-world communication of residues and significance for protein dynamics. *Biophys J* 86(1 Pt 1):85-91.
78. del Sol A, Tsai CJ, Ma B, & Nussinov R (2009) The origin of allosteric functional modulation: multiple pre-existing pathways. *Structure* 17(8):1042-1050.
79. Tsai CJ & Nussinov R (2014) A Unified View of "How Allostery Works". *Plos Comput Biol* 10(2).
80. Kar G, Keskin O, Gursoy A, & Nussinov R (2010) Allostery and population shift in drug discovery. *Curr Opin Pharmacol* 10(6):715-722.
81. Reichardt J & Bornholdt S (2006) Statistical mechanics of community detection. *Phys Rev E* 74(1).
82. Traag VA & Bruggeman J (2009) Community detection in networks with positive and negative links. *Phys Rev E* 80(3).

83. Malik G, Banerji A, & Kloczkowski A (2017) Deciphering General Characteristics of Residues Constituting Allosteric Communication Paths. *Biophys J* 112(3):499a-499a.
84. Malik G & Kloczkowski A (2018) Classification of Allostery in Proteins: A Deep Learning Approach. *Biophysical Journal* 114(3):422a %@ 0006-3495.
85. Grindley HM, Artymiuk PJ, Rice DW, & Willett P (1993) Identification of Tertiary Structure Resemblance in Proteins Using a Maximal Common Subgraph Isomorphism Algorithm. *Journal of Molecular Biology* 229(3):707-721.
86. Koch I, Lengauer T, & Wanke E (1996) An algorithm for finding maximal common subtopologies in a set of protein structures. *J Comput Biol* 3(2):289-306.
87. Raymond JW & Willett P (2002) Maximum common subgraph isomorphism algorithms for the matching of chemical structures. *J Comput Aid Mol Des* 16(7):521-533.
88. Koch I (2001) Enumerating all connected maximal common subgraphs in two graphs. *Theor Comput Sci* 250(1-2):1-30.
89. Mcgregor JJ (1982) Backtrack Search Algorithms and the Maximal Common Subgraph Problem. *Software Pract Exper* 12(1):23-34.
90. Barabasi AL & Albert R (1999) Emergence of scaling in random networks. *Science* 286(5439):509-512.
91. Barabasi AL & Oltvai ZN (2004) Network biology: Understanding the cell's functional organization. *Nat Rev Genet* 5(2):101-U115.
92. Watts DJ & Strogatz SH (1998) Collective dynamics of 'small-world' networks. *Nature* 393(6684):440-442.
93. Wuchty S (2001) Scale-free behavior in protein domain networks. *Mol Biol Evol* 18(9):1694-1702.
94. Vendruscolo M, Dokholyan NV, Paci E, & Karplus M (2002) Small-world view of the amino acids that play a key role in protein folding. *Phys Rev E* 65(6).
95. Greene LH & Higman VA (2003) Uncovering network systems within protein structures. *Journal of Molecular Biology* 334(4):781-791.
96. Barahona M & Pecora LM (2002) Synchronization in small-world systems. *Phys Rev Lett* 89(5).
97. Nishikawa T, Motter AE, Lai YC, & Hoppensteadt FC (2003) Heterogeneity in oscillator networks: Are smaller worlds easier to synchronize? *Phys Rev Lett* 91(1).
98. Roxin A, Riecke H, & Solla SA (2004) Self-sustained activity in a small-world network of excitable neurons. *Phys Rev Lett* 92(19).
99. Lago-Fernandez LF, Huerta R, Corbacho F, & Siguenza JA (2000) Fast response and temporal coherent oscillations in small-world networks. *Phys Rev Lett* 84(12):2758-2761.
100. del Sol A & O'Meara P (2005) Small-world network approach to identify key residues in protein-protein interaction. *Proteins* 58(3):672-682.
101. Daily MD, Upadhyaya TJ, & Gray JJ (2008) Contact rearrangements form coupled networks from local motions in allosteric proteins. *Proteins* 71(1):455-466.
102. Schellman JA (1959) The Factors Affecting the Stability of Hydrogen-Bonded Polypeptide Structures in Solution. *J Phys Chem-Us* 62(12):1485-1494.

103. Kim H, *et al*. (2015) A Hinge Migration Mechanism Unlocks the Evolution of Green-to-Red Photoconversion in GFP-like Proteins. *Structure* 23(1):34-43.
104. Na H, Lin TL, & Song G (2014) Generalized Spring Tensor Models for Protein Fluctuation Dynamics and Conformation Changes. *Adv Exp Med Biol* 805:107-135.
105. Song G & Jernigan RL (2006) An enhanced elastic network model to represent the motions of domain-swapped proteins. *Proteins* 63(1):197-209.
106. Jamroz M, Kolinski A, & Kihara D (2012) Structural features that predict real-value fluctuations of globular proteins. *Proteins* 80(5):1425-1435.
107. Yang YD, Park C, & Kihara D (2008) Threading without optimizing weighting factors for scoring function. *Proteins* 73(3):581-596.
108. Enright MB & Leitner DM (2005) Mass fractal dimension and the compactness of proteins. *Phys Rev E* 71(1).
109. Banerji A & Ghosh I (2009) Revisiting the Myths of Protein Interior: Studying Proteins with Mass-Fractal Hydrophobicity-Fractal and Polarizability-Fractal Dimensions. *Plos One* 4(10).
110. Leitner DM (2008) Energy flow in proteins. *Annu Rev Phys Chem* 59:233-259.
111. Reuveni S, Granek R, & Klafter J (2010) Anomalies in the vibrational dynamics of proteins are a consequence of fractal-like structure. *P Natl Acad Sci USA* 107(31):13696-13700.

**Figure Labels/Titles and Figure Legends :**

Figure 1 Title: Cliques and Communities in ACSS networks.

Figure-1 Legend: Cliques and communities found in 30 ACSS under study. Residues are identified in the format [Residue-Number in a protein:The coarse-grained character of the residue]. The coarse-grained character labelling scheme employed is: Acidic-A, Basic-B, Cysteine-C, Proline-P, Hydrophobic-H, Hydroxyl-X, Amide-E, Small-S.

Figure 2 Title: Maximal-Common-Subgraph generated from different subsets of Kinase ACSS.

Figure-2 Legend: The Maximal-Common-Subgraph (viz., maximal common pattern present in the ACSS cliques) of different subsets of Kinase ACSS. Fig 2.1. demonstrates 'Hydrophobic-Hydrophobic-Basic' embodied among PDB ids:1CZA, 1PFK, 1SNI, 2BTZ, 2XRW ACSS cliques; Fig 2.2 demonstrates 'Hydroxyl-Hydrophobic-Basic' embodied among PDB id.s:1DKU, 1PFK, 3MK6, 4AW0 ACSS cliques; Fig2.3 demonstrates 'Hydrophobic-Acidic-Hydrophobic' embodied among PDB ids:1E0T, 1PFK, 1SQ5, 2BTZ ACSS cliques; Fig 2.4. demonstrates 'Hydrophobic-Basic-Basic' embodied among PDB id.s:1PFK, 2JJX, 2VTT ACSS cliques; Fig 2.5. demonstrates 'Hydrophobic-Hydrophobic-Hydrophobic' embodied among PDB id.s:1SNI, 1SQ5, 2BTZ, 3BQC, 3EQC ACSS cliques. Details of ACSS cliques can be found from Fig.-1 and from Supplementary Material-3. The coarse-grained character labelling scheme employed is: Acidic-A, Basic-B, Cysteine-C, Proline-P, Hydrophobic-H, Hydroxyl-X, Amide-E, Small-S.

Figure-3: The SWN of PDB Id.-1XNX's ACSS.

Figure-3 Legend: The entire scope of ACSS of 1XNX was found to generate a SWN. Details of the SWN can be found in Table-4.A. The coarse-grained character labelling scheme employed is: Acidic-A, Basic-B, Cysteine-C, Proline-P, Hydrophobic-H, Hydroxyl-X, Amide-E, Small-S.

Figure4: Comparison of RMSF values from the MD simulations and B-factor values from X-ray crystallography.

Figure-4 Legend: Comparison of RMSF values from the MD simulations and B-factor values from X-ray crystallography for three proteins identified by PDB codes 1Q5Y (a), 1SC3 (b) and 2JXX (c). Top and bottom snapshots show protein's cartoon representation coloured by B-factor and RMSF values, respectively. Note that conformations shown at bottom represents average conformation of most probably cluster from the CABS-flex simulation.

**Table 1: B-Factor of residues constituting ACSS and those constituting non-ACSS**

| Kin. ACSS | | Kin. non-ACSS | | Pept. ACSS | | Pept. non-ACSS | | N.R. ACSS | | N.R. non-ACSS | | T.F. ACSS | | T.F. non-ACSS | |
|---|---|---|---|---|---|---|---|---|---|---|---|---|---|---|---|
| Res. | B-factor | Res. | B-factor | Res. | B-factor | Res. | B-factor | Res. | B-factor | Res. | B-factor | Res. | B-factor | Res. | B-fact. |
| ALA | 36.53(24.21) | ALA | 30.34(17.73) | | | ALA | 20.89(14.36) | | | ALA | 32.89(22.47) | ALA | 25.94(4.15) | ALA | 37.71(19.94) |
| ARG | 35.21(21.93) | ARG | 37.69(21.34) | ARG | 25.69(13.14) | ARG | 23.68(16.86) | ARG | 27.79(17.02) | ARG | 46.01(24.03) | ARG | 46.38(17.76) | ARG | 46.82(22.25) |
| ASN | 59.04(30.24) | ASN | 36.84(19.79) | | | ASN | 24.04(16.16) | ASN | 13.86(5.30) | ASN | 44.61(23.59) | ASN | 28.83(19.51) | ASN | 52.78(24.03) |
| ASP | 35.64(13.31) | ASP | 38.09(21.57) | ASP | 14.62(3.92) | ASP | 26.64(19.43) | | | ASP | 40.96(24.86) | ASP | 38.26(12.27) | ASP | 42.54(21.98) |
| | | CYS | 29.03(16.94) | CYS | 35.69(1.38) | CYS | 20.16(14.56) | | | CYS | 38.71(27.67) | CYS | 52.52(26.14) | CYS | 28.70(13.13) |
| GLN | 28.44(9.25) | GLN | 39.92(22.96) | | | GLN | 25.74(21.61) | GLN | 54.70(5.09) | GLN | 35.91(20.80) | GLN | 36.90(4.65) | GLN | 39.24(20.99) |
| GLU | 39.59(17.93) | GLU | 38.99(20.70) | GLU | 15.95(4.99) | GLU | 32.80(18.88) | GLU | 51.66(18.04) | GLU | 47.97(26.94) | | | GLU | 53.41(24.19) |
| GLY | 27.87(9.56) | GLY | 32.86(17.22) | | | GLY | 22.80(16.68) | | | GLY | 47.41(27.43) | | | GLY | 39.62(22.36) |
| HIS | 44.97(20.35) | HIS | 35.71(20.97) | | | HIS | 27.35(20.58) | HIS | 13.88(2.28) | HIS | 37.59(23.26) | HIS | 37.62(18.24) | HIS | 34.54(21.18) |
| ILE | 42.26(16.44) | ILE | 29.78(15.62) | | | ILE | 21.97(15.70) | | | ILE | 44.0(28.05) | | | ILE | 38.45(19.11) |
| LEU | 40.69(19.53) | LEU | 29.97(16.05) | | | LEU | 19.34(13.89) | LEU | 54.75(5.83) | LEU | 34.14(23.81) | | | LEU | 40.07(20.85) |
| LYS | 38.72(11.67) | LYS | 38.81(20.22) | | | LYS | 31.09(18.67) | | | LYS | 46.38(27.14) | LYS | 47.32(3.57) | LYS | 53.42(23.40) |
| MET | 43.12(22.56) | MET | 31.10(15.56) | | | MET | 18.20(14.14) | MET | 59.16(0.80) | MET | 41.21(24.96) | | | MET | 42.64(23.60) |
| PHE | 28.48(6.08) | PHE | 29.68(16.62) | | | PHE | 22.49(15.94) | PHE | 16.25(2.38) | PHE | 37.18(23.42) | | | PHE | 41.84(23.56) |
| | | PRO | 33.25(17.87) | | | PRO | 21.53(16.09) | | | PRO | 39.67(25.50) | | | PRO | 41.43(19.17) |
| SER | 34.86(13.03) | SER | 33.41(19.38) | SER | 11.33(0.60) | SER | 23.36(17.02) | SER | 8.37(0.82) | SER | 38.91(26.37) | SER | 17.47(1.97) | SER | 41.46(21.70) |
| THR | 45.03(27.45) | THR | 32.29(18.18) | THR | 12.35(0.40) | THR | 20.91(15.51) | | | THR | 42.18(25.24) | THR | 51.90(20.00) | THR | 40.01(19.26) |
| | | TRP | 31.22(23.75) | TYR | 36.82(1.73) | TRP | 27.45(22.24) | | | TRP | 44.96(32.35) | | | TRP | 32.68(13.28) |
| TYR | 52.48(24.13) | TYR | 31.12(16.42) | | | TYR | 22.70(18.29) | TYR | 11.47(3.65) | TYR | 43.32(28.58) | TYR | 67.23(2.49) | TYR | 39.88(18.00) |
| VAL | 38.10(23.12) | VAL | 28.78(14.69) | | | VAL | 19.41(15.28) | VAL | 57.18(0.37) | VAL | 38.22(24.76) | | | VAL | 37.22(21.24) |

**Table 1 legend:** B-factors were calculated at the residual level in ACSS and non-ACSS, they are presented as Mean(Std. Dev.)

**Table 2: The Centrality Indices for ACSS and non-ACSS for four groups of allosteric proteins.**

|  | Kinase ACSS Residues | Kinase non-ACSS Residues | Peptidase ACSS Residues | Peptidase non-ACSS Residues | Nuclear receptor ACSS Residues | Nuclear receptor non-ACSS Residues | Transcription Factor ACSS Residues | Transcription Factor non-ACSS Residues |
|---|---|---|---|---|---|---|---|---|
| **Average Degree Centrality** | 0.411 | 0.518 | 0.722 | 0.814 | 0.613 | 0.690 | 0.426 | 0.659 |
| **Average Closeness Centrality** | 0.488 | 0.596 | 0.809 | 0.843 | 0.707 | 0.754 | 0.474 | 0.719 |
| **Average Betweenness Centrality** | 0.075 | 0.122 | 0.194 | 0.038 | 0.137 | 0.141 | 0.062 | 0.119 |

**Table 2 Legend:** The three types of major centrality measures calculated on the ACSS and non-ACSS graphs of the same size.

**Table 3: Comparison of Global Clustering Coefficient Average and Global Average Shortest Path Length for ACSS and non-ACSS**

|  | Kinases | | Nuclear Receptors | | Peptidases | | Transcription Factors | |
| --- | --- | --- | --- | --- | --- | --- | --- | --- |
|  | ACSS | Non-ACSS | ACSS | Non-ACSS | ACSS | Non-ACSS | ACSS | Non-ACSS |
| **Global Clustering Coefficient Average** | 0.145 | 0.559 | 0.208 | 0.594 | 0.0 | 0.521 | 0.0 | 0.584 |
| **Global Average Shortest Path Length** | 1.229 | 7.396 | 0.976 | 6.587 | 1.0 | 6.136 | 1.119 | 5.042 |

**Table: 4.A.**

| Group | Protein | Is it a Small world for Erdös-Rényi random graph | | | Nodes in Graph | Number of Nodes | Number of Edges | Average Clustering Coefficient | Average Shortest Path Length |
|---|---|---|---|---|---|---|---|---|---|
| | | P=0.3 | P=0.5 | P=0.7 | | | | | |
| Nuclear Receptors | 1XNX | N | Y | Y | ['A', 'H', 'E', 'B', 'H', 'A', 'A'] | 7 | 16 | 0.838 | 1.238 |

**Table 4.A legend: Summary of SWN characteristics for ACSS residues of 1XNX.**

**Table 4.B.**

| Group | Protein | Is it a Small world for Erdös-Rényi random graph | | | Nodes in Graph | Number of Nodes | Number of Edges | Average Clustering Coefficient | Average Shortest Path Length |
|---|---|---|---|---|---|---|---|---|---|
| | | P=0.3 | P=0.5 | P=0.7 | | | | | |
| Kinases | 2BTZ _subset | Y | Y | N | ['H', 'X', 'E', 'H', 'H', 'H', 'H'] | 7 | 10 | 0.59 | 1.619 |
| | 2JJX(A)_subset_1 | Y | Y | N | ['E', 'S', 'A'] | 3 | 3 | 1 | 1 |
| | 2JJX(A)_subset_2 | Y | Y | N | ['B', 'H', 'B'] | 3 | 2 | 0 | 1.333 |
| | 3EQC_subset | Y | Y | N | ['H', 'H', 'X', 'H', 'H', 'H', 'H'] | 7 | 8 | 0.571 | 2.048 |
| | 3MK6(A)_subset | Y | N | N | ['X', 'H', 'B'] | 3 | 2 | 0 | 1.333 |
| Nuclear Receptors | 2AX6_subset | N | Y | N | ['E', 'H', 'A'] | 3 | 2 | 0 | 1.333 |
| Peptidases | No SWN was observed even among the subgraphs of residues constituting ACSSs of peptidases. | | | | | | | | |
| Transcription Factors | No SWN was observed among the subgraphs of residues constituting ACSSs of transcription factors. | | | | | | | | |

**Table 4.B legend: Summary of SWN characteristics in the subgraphs of ACSSs**

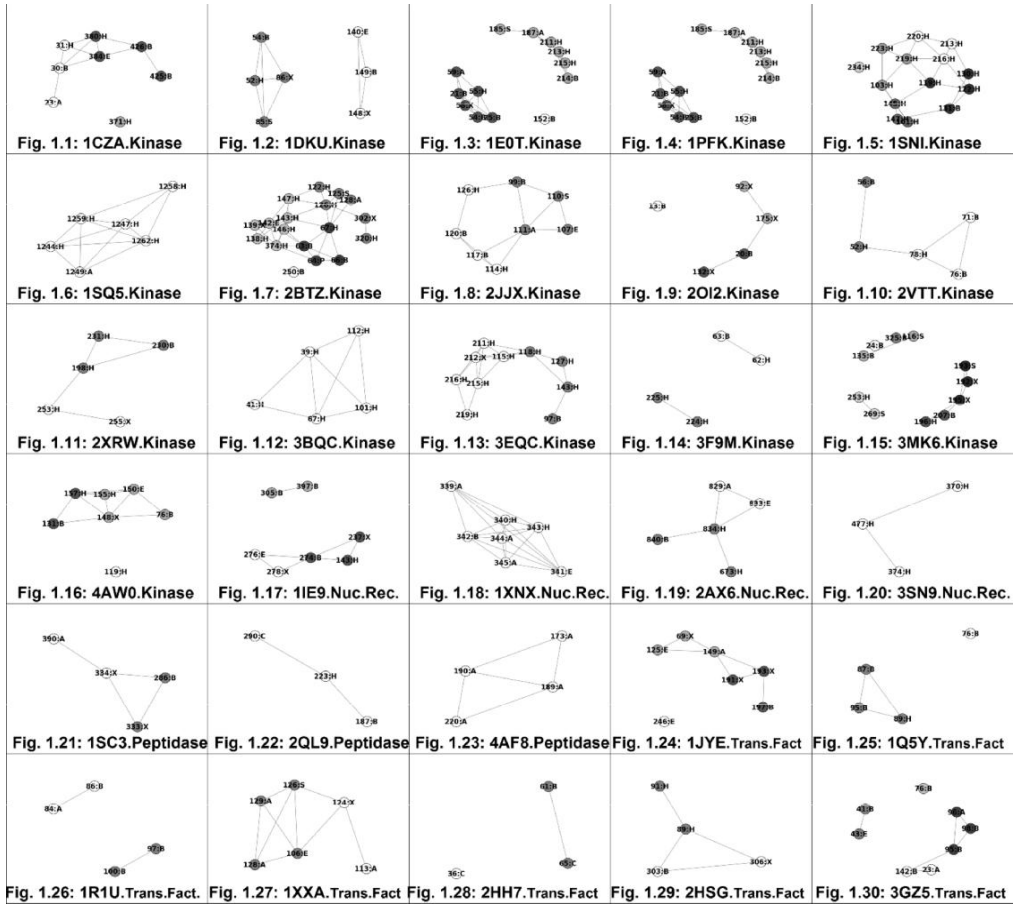

**Figure 1.** Cliques and Community Graphs

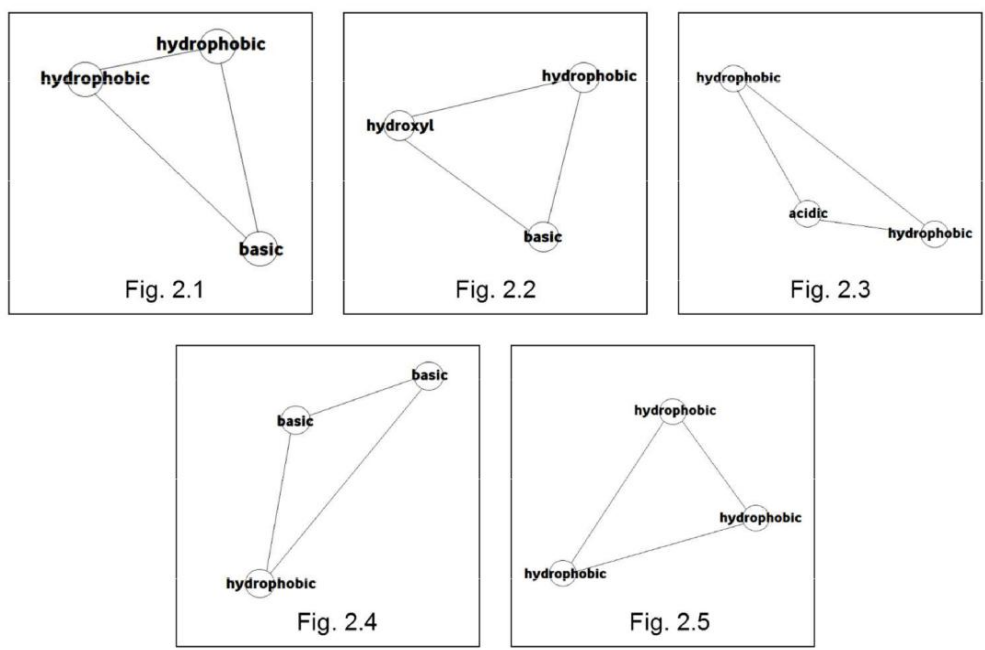

**Figure 2.** Maximum Common Subgraphs

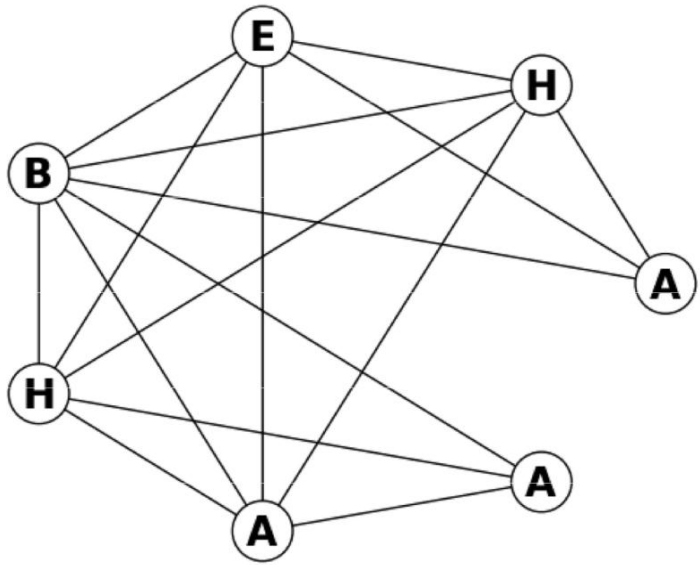

**Figure 3.** *Small World Network*

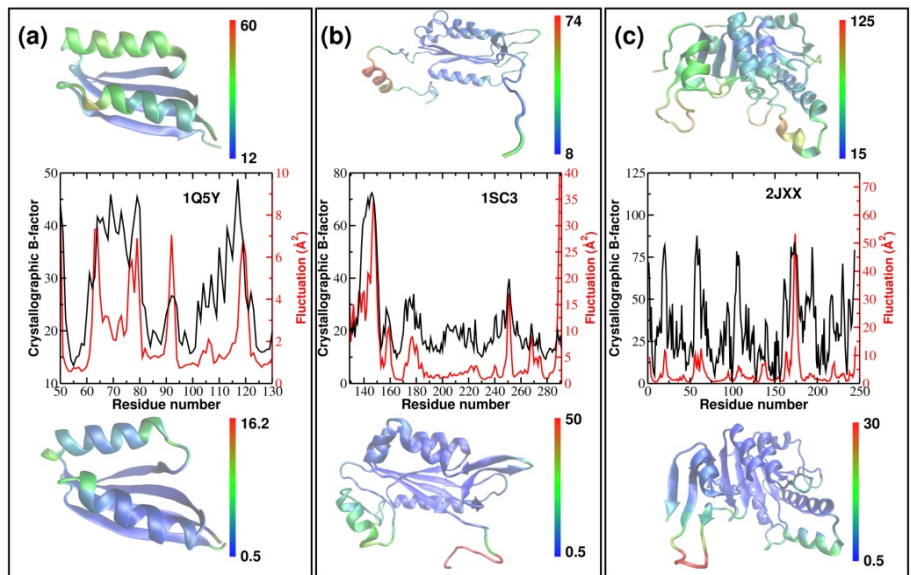

**Figure 4.** Comparison of RMSF Values with B factors